\documentclass[letterpaper,aps,pra,showpacs,twocolumn,floatfix,nofootinbib,amsmath,
amssymb]{revtex4-1}
\usepackage{color,graphicx}
\usepackage{dcolumn}
\usepackage{bm}
\usepackage[normalem]{ulem}
\makeatletter

\newcommand{\Rmnum}[1]{\expandafter\@slowromancap\romannumeral #1@}
\makeatother

\begin{document}

\title{A new effective interaction for the trapped fermi gas: the BEC-BCS crossover}

\author{C. N. Gilbreth and Y. Alhassid}

\affiliation{Center for Theoretical Physics, Sloane Physics
 Laboratory, Yale University, New Haven, CT 06520}

\begin{abstract}
  We extend a recently introduced separable interaction for the unitary trapped
  Fermi gas to all values of the scattering length. We derive closed expressions
  for the interaction matrix elements and the two-particle eigenvectors and
  analytically demonstrate the convergence of this interaction to the zero-range
  two-body pseudopotential for $s$-wave scattering. We apply this effective
  interaction to the three- and four-particle systems along the BEC-BCS
  crossover, and find that their low-lying energies exhibit convergence in the
  regularization parameter that is much faster than for the conventional
  renormalized contact interaction. We find similar convergence properties of
  the three-particle free energy at unitarity.
\end{abstract}


\maketitle

\section{Introduction}

Atomic Fermi gas systems have generated much interest in the last several
years~\cite{giorgini:1215,bloch:885}. While these systems can be well-described
by a relatively simple model Hamiltonian, at low temperatures they exhibit a
rich phenomenology whose theoretical description has proven to be challenging.
At low momenta, or when the range of the interaction is sufficiently short, the
dominant scattering process occurs in the $s$-wave channel, so that a single
parameter, the $s$-wave scattering length $a$, suffices to characterize the
inter-particle interaction. Depending on the value of $k_F a$ (where $k_F$ is
the Fermi momentum), widely different behavior is observed. This ranges from the
formation of tightly bound dimers in the Bose-Einstein condensate (BEC) regime
when $k_F a$ is small and positive, to Cooper pairing in the
Bardeen-Cooper-Schrieffer (BCS) regime when $k_F a$ is small and negative. These
behaviors are connected by a strongly interacting, nonperturbative regime, known
as the unitary regime, where $|a|$ is much larger than any other length scale in
the system. Remarkably, each of these regimes is accessible experimentally, and
all exhibit superfluid behavior below a certain $a$-dependent critical
temperature.

While accurate mean-field theories exist for the BEC and BCS regimes, there is
no simple approximation that can accurately describe the transition between
these regimes through the unitary limit. As a result much interest has been
taken in applying numerical methods, such as quantum Monte Carlo and numerical
diagonalization. For systems varying from $\sim 10$ to hundreds of particles,
quantum Monte Carlo calculations have been carried out to calculate ground
state~\cite{carlson:2003,astrakharchik:2004,chang:2004,chang:2005,stecher:2007,
  blume:2007,jauregui:2007,greene:2008,gezerlis:2008,zinner:2009} and
thermodynamic properties~\cite{bulgac:2006-9,burovski:2006,chang:2007,akkineni:2007,goulko:2010}.
For smaller numbers of trapped atoms, energy spectra have been calculated using
a basis set expansion method with correlated Gaussians in coordinate space (for
up to six atoms)~\cite{stecher:2007,blume:2007} and a stochastic variational
approach (for four atoms)~\cite{daily:2010}.

The three-body problem has been solved analytically in the unitary
regime~\cite{werner:2006,tan3body} and numerically to high accuracy along the
BEC-BCS crossover~\cite{Kestner:2007,Liu:2009}. In fact the few-body trapped
cold atom problem has become more interesting recently, since it was pointed out
that the virial expansion for the partition function in a harmonic trap works
well at unexpectedly low temperatures into the quantum degenerate
regime~\cite{Liu:2009}. Moreover, the scaling relations which exist for the
dilute gas can allow larger systems to be addressed from the study of smaller
ones~\cite{adhikari:2009,zhang:2009}.

Here we discuss the cold atom problem in the context of the
configuration-interaction (CI) approach to interacting
many-particle fermionic systems. In this approach, widely used in atomic,
molecular and nuclear physics, a single-particle basis in the laboratory frame
is used to construct a many-particle basis of Slater determinants for a fixed
number of fermions. The Hamiltonian matrix is then calculated in this
many-particle basis and diagonalized. For a harmonically trapped system, a
natural choice for the single-particle basis is that of the three-dimensional
harmonic oscillator.

The short-range interaction of the cold atom problem is often approximated by a
zero-range interaction. However, the obvious form of such a potential, a contact
interaction $V_0 \delta(\bm{r})$, is ill-defined in three dimensions and must be
regularized. This is usually accomplished by introducing a momentum or energy
cutoff in relative motion and renormalizing the strength $V_0$ of the
interaction so it reproduces the physical scattering length for the uniform
gas, or the lowest bound state energy in relative motion for the trapped
system. The many-particle energies are then calculated as a function of the
regularization cutoff parameter.

A contact interaction is non-vanishing only for relative angular momenta $l = 0$
of the two particles. For a basis of harmonic oscillator wave functions, a
natural regularization parameter is the number of oscillator $s$ waves in
relative motion. However, the convergence of the many-particle energies as a
function of this regularization parameter for the renormalized contact
interaction is slow (as a low negative power)~\cite{Stetcu:2008, Alhassid:2008}.

In Ref.~\onlinecite{Alhassid:2008} a new effective interaction was introduced in
the unitary limit for which the low-lying energies of the three- and
four-particle systems converge substantially faster than for the conventional
renormalized contact interaction as a function of the regularization parameter.
While this effective interaction is no longer a contact interaction, it
reproduces the same many-particle energies in the limit when the regularization
parameter is sent to infinity. The faster convergence of the many-particle
energies enables their calculation to higher accuracy when using this new
effective interaction. Another approach to improve the convergence relative to
the renormalized contact interaction was investigated within the framework of
effective field theory by including perturbatively next-to-leading order and
next-to-next-to-leading order interactions~\cite{Stetcu:2010}.

Here we generalize the construction of such an effective interaction away from
unitarity. We study the low-energy spectra of the three- and four-particle
systems and find convergence properties versus the regularization parameter along
the complete BEC-BCS crossover that are similar to the unitary regime. We also
use this effective interaction to study the thermodynamics of the three-particle
system at unitarity. In particular, we demonstrate the exponential convergence
of the free energy at finite temperature as a function of the regularization
parameter. The converged free energy is compared with the exact free energy
constructed from the known analytical spectrum of the three-particle
system~\cite{werner:2006,tan3body}.

This paper is organized as follows. In Sec.~\Rmnum{2}, we briefly review the
two-particle cold atom problem in a harmonic trap. In Sec.~\Rmnum{3}, we derive
closed expressions for the matrix elements of the new effective interaction for
any scattering length and demonstrate convergence (as a function of the
regularization parameter) of its two-particle eigentates to the exact
eigenstates in the unitary limit. In Sec.~\Rmnum{4}, we use this interaction to
study the convergence (vs. the regularization parameter) of the spectra of the
three- and four-body systems along the BEC-BCS crossover and the three-particle
free energy at unitarity. The latter is compared with exact results. Finally, in
Sec.~\Rmnum{5} we present our conclusion.

\section{Two-particle problem}

The trapped two species cold atom system is modeled by the Hamiltonian
\begin{equation}
  H = - \sum_{i = 1}^{N} \frac{\hbar^2}{2 m} \nabla_i^2 + \sum_{i =
    1}^{N}
  \frac{1}{2} m \omega^2 r_i^2 + \sum_{i < j} V (\bm{r}_{i j}),
  \label{eq:H1}
\end{equation}
where $N=N_1+N_2$, $N_1$ and $N_2$ are the number of atoms for each species
(spin-up and spin-down), $\omega$ is the trap frequency, and $V (\bm{r})$ is a
short-range interaction. A natural length scale for this Hamiltonian is the
oscillator length $a_{\textrm{osc}}=\sqrt{\hbar/m\omega}$.

The interaction $V (\bm{r})$ is modeled by a zero-range,
pure $s$-wave interaction, which due to Pauli exclusion acts only between
particles of differing species. This can be expressed by a regularized
$\delta$-function,
\begin{equation}
  V (r) = g \delta^3 (\bm{r}) (\partial /\partial r) r \;, \label{eq:pseudo}
\end{equation}
where $g = 2 \pi \hbar^2 a / \mu$ relates the scattering length $a$ to the
strength of the interaction \cite{giorgini:1215}. Equivalently, one can impose
the Bethe-Peierls contact conditions on the wavefunctions; see, e.g., in
Ref.~\onlinecite{werner:2006}. Several interesting consequences of the form
(\ref{eq:pseudo}) of the interaction potential for the many-body system have
been explored in Ref.~\onlinecite{tan:2008}.

For a two-particle system, the Hamiltonian is separable in the center-of-mass
and relative coordinates $\bm{R}=(\bm{r}_1 + \bm{r}_2)/2$ and $\bm{r}=\bm{r}_2
-\bm{r}_1$, so that $H = H_{\rm CM} + H_{\rm rel}$, with
\[ H_{\rm rel} = -\frac{\hbar^2}{2 \mu} \nabla_{\bm{r}}^2 + \frac{1}{2} \mu
\omega^2 r^2 + V (\bm{r}) \equiv H_0 +V(r) \;,\] where $H_0$ is the harmonic
oscillator Hamiltonian in relative motion with reduced mass $\mu = m / 2$.

The eigenstates of the non-interacting two-particle system in a harmonic trap
may be labeled as $|\mathcal{N L M} n l m\rangle$, where $\mathcal{N,L,M}$ are
the radial, angular momentum, and magnetic quantum numbers for the center of
mass motion, and $n,l,m$ are the corresponding quantum numbers for the relative
motion. The associated energies are $E = (2\mathcal{N}+\mathcal{L}+ 3 / 2 + 2n +
l + 3/2)\hbar \omega$. A pure $s$-wave interaction leaves the $l \ne 0$ states
and energies unchanged while mixing the $l=0$ states into eigenstates we denote
by $|\mathcal{N L M} u^{(i)}\rangle$, with $E = 2\mathcal{N}+\mathcal{L}+ 3 / 2
+ \varepsilon_i$. The energies $\varepsilon_{i}$ in relative motion for
scattering length $a$ are given by the solutions to the transcendental
equation~\cite{Busch:1998,Jonsell:2002}
\begin{equation}
  \frac{\Gamma (- \varepsilon / (2\hbar \omega) + 3 / 4)}{\Gamma (- \varepsilon / (2\hbar \omega) + 1 /
    4)} = \frac{1}{\sqrt{2}a/a_{\rm osc}} . \label{eq:energies}
\end{equation}

The exact wave functions in relative motion are also known. They
are given by \cite{Busch:1998}
\[ | u^{(i)} \rangle = \sum_{n = 0}^{\infty} u^{(i)}_n |n 0 0 \rangle \;, \]
where
\[ u^{(i)}_n = A_i \frac{\varphi_n^{\ast} (0)}{\alpha_n - \varepsilon_i} \;. \]
Here $\alpha_n=(2n+3/2)\hbar \omega$ are the non-interacting relative energies,
$A_i$ is a normalization factor, and $\varphi_n (\bm{r}) \equiv \varphi_{n 0 0}
(\bm{r})$ is a harmonic oscillator wave function, with
\[ \varphi_{n} (0) = \pi^{- 3 / 4} \left[ \frac{(2 n + 1) !!}{(2 n) !!}
\right]^{1 / 2}\; . \]

In the unitary limit, where $\varepsilon_i=(2i+1/2)\hbar\omega$, the
normalization factor $A_i$ has the simple form
\begin{eqnarray*}
  A_i^{- 2} & = & \pi^{- 3 / 2} \sum_{n = 0}^{\infty} \frac{(2 n + 1) !!}{(2 n)
    !!}  \frac{1}{[2 (n - i) + 1]^2} \label{eq:A2} \\
  & = & {\pi^{- 1 / 2} \over 2}  \frac{(2 i) !!}{(2 i - 1) !!} \;.
\end{eqnarray*}

\section{Effective interaction}

We first discuss the conventional contact interaction. It can be regularized by
introducing an energy cutoff in relative motion. This defines a sequence of
interactions $\hat{V}_{\rm c}^{(q)}$, $q=0,1,2,...$, with
\begin{equation}
  \langle n l m| \hat{V}_{\rm c}^{(q)} |n' l m \rangle = \chi_q \psi_n (0) \psi_{n'} (0)
  \delta_{l, 0} \delta_{m, 0} \;, \label{eq:contact}
\end{equation}
for $n,n'=0,1,\ldots,q$, and where $\psi_{n}(0) \equiv [(2n+1)!!/(2n)!!]^{1/2}$.
All other matrix elements of $V_{\rm c}^{(q)}$ are taken to be zero. The
coefficient $\chi_q$ is determined for each $q$ so as to reproduce the
lowest relative-motion energy $\varepsilon_0$. We refer to this interaction as a
renormalized contact interaction.

The contact interaction in Eq.~(\ref{eq:contact}) is separable in the relative
oscillator basis. A general separable interaction for $l=0$ has the form
\begin{equation}
  \langle n l m| \hat{V}_{\rm eff}^{(q)} |n' l m \rangle = f^*_n f_{n'} \delta_{l, 0}
  \delta_{m, 0} \hspace{2em} (n, n' \leqslant q) \;. \label{eq:effective}
\end{equation}
An effective interaction for the trapped cold atom system can be determined by
choosing the coefficients $f_n$ to reproduce the lowest $q+1$ relative energies
$\varepsilon_0, \ldots, \varepsilon_q$ of the two-particle system. This idea was
used in Ref.~\onlinecite{Alhassid:2008} to construct a new effective interaction
in the unitary limit of infinite scattering length. Here we consider the more
general problem for any value of the scattering length $a$.

\subsection{General problem and solution}

The matrix elements of the relative-coordinate Hamiltonian take the form
\begin{equation}
  \langle n 0 0| \hat{H}^{(q)}_{\rm rel} |n' 0 0 \rangle = \delta_{n, n'} \alpha_n -
  f^*_n f_{n'} \;,
\end{equation}
where $\alpha_n$ are the non-interacting eigenvalues of $H_0$. To determine the
coefficients $f_n$ we derive the following result.

\emph{Theorem.} Let $\varepsilon_0, \varepsilon_1, \ldots,\varepsilon_q$ and
$\alpha_0, \alpha_1, \ldots \alpha_q$ be real numbers such that $\varepsilon_0 <
\ldots < \varepsilon_{q}$, $ \alpha_0 < \ldots < \alpha_q$ and $\varepsilon_i
\neq \alpha_j$ for all $i,j$, and let $f_0,f_1,\ldots,f_q$ be complex. Then the
$(q+1)$-dimensional matrix $H_{n,n'} = \alpha_n \delta_{n,n'} - f_n^{*} f_{n'}$
has the $\varepsilon_i$ as its eigenvalues if and only if
\begin{equation}
  |f_n| = \sqrt{\frac{\prod_{k} (\alpha_n - \varepsilon_k)} {\prod_{k\neq
        n}(\alpha_n-\alpha_k)}},
  \label{absfn}
\end{equation}
in which case the $i$th eigenvector $b^{(i)}$ has components
\begin{equation}
  b_n^{(i)} = C_i \frac{f_n^{*}}{\alpha_n - \varepsilon_i}, \qquad
  C_i=\sqrt{\frac{\prod_k(\alpha_k-\varepsilon_i)}{\prod_{k \neq i}(\varepsilon_k-\varepsilon_i)}}
  \; .
\end{equation}
A solution satisfying (\ref{absfn}) exists if and only if $\varepsilon_0 <
\alpha_0 < \varepsilon_1 < \ldots < \alpha_q$.

\emph{Proof.}
An eigenvector $b^{(i)}$ with eigenvalue $\varepsilon_i$ must satisfy
\begin{equation}
  \sum_{n=0}^q (\delta_{m n} \alpha_n - f_m^* f_n) b^{(i)}_n = \varepsilon_i b^{(i)}_m
  \label{eq:eigenvalue}
\end{equation}
which implies
\begin{equation}
  b^{(i)}_n = C_i \frac{f_n^*}{\alpha_n - \varepsilon_i}
  \label{eq:bj}
\end{equation}
where $C_i = \sum_{m} f_m b_m^{(i)}$ is a normalization constant. Inserting
(\ref{eq:bj}) into (\ref{eq:eigenvalue}), we obtain
\begin{equation*}
  \sum_{n=0}^q M_{i n} |f_n|^2 = 1 \;, \label{eq:findfj}
\end{equation*}
where the matrix $M$ is defined by
\begin{equation}
  M_{i n} = \frac{1}{\alpha_n - \varepsilon_i} \;.
\end{equation}
Thus the $f_n^2$ are determined from the eigenvalues by the matrix equation $M
\bm{v} =\bm{1}$, where $\bm{1}= (1, 1, \ldots, 1)^T$ and $\bm{v} = (|f_0|^2,
|f_1|^2, \ldots, |f_q|^2)^T$.

The Matrix $M$ is of a well-known form, called a Cauchy matrix, whose
inverse is given by~\cite{knuthv1}
\[ (M^{- 1})_{n i} = \frac{\prod_k (\alpha_n - \varepsilon_k) (\alpha_k -
  \varepsilon_i)}{(\alpha_n - \varepsilon_i) \prod_{k \neq i} (\varepsilon_k
  - \varepsilon_i) \prod_{k \neq n} (\alpha_n - \alpha_k)} . \]
We therefore have
\begin{eqnarray}
  |f_n|^2 & = & (M^{- 1} \bm{1})_n \nonumber \\
  & = & \sum_i \frac{\prod_k (\alpha_n -
    \varepsilon_k) (\alpha_k - \varepsilon_i)}{(\alpha_n - \varepsilon_i)
    \prod_{k \neq i} (\varepsilon_k - \varepsilon_i) \prod_{k \neq n} (\alpha_n
    - \alpha_k)} \nonumber \\
  & = & \frac{\prod_k (\alpha_n - \varepsilon_k)}{\prod_{k \neq n} (\alpha_n
    - \alpha_k)} \sum_i \frac{\prod_{k \neq n} (\varepsilon_i -
    \alpha_k)}{\prod_{k \neq i} (\varepsilon_i - \varepsilon_k)} \;. \label{f^2}
\end{eqnarray}
The sum in the last expression can be evaluated using the identity \cite{knuthv1}
\begin{equation}
  \sum_{i = 1}^N \frac{x_i^r}{\prod_{k \neq i} (x_i - x_k)} = \left\{
    \begin{array}{ll}
      0, & 0 \leqslant r < N - 1 \\
      1, & r = N - 1 \\
      \sum_{i = 1}^N x_i, & r = N
    \end{array} \label{eq:identity} \right.
\end{equation}
where $x_1, \ldots, x_N$ are all distinct. This is done by observing that
$\prod_{k \neq n} (\varepsilon_i - \alpha_k) = \varepsilon^{q}_i + P
(\varepsilon_i)$, where $P$ is a polynomial of degree smaller than $q$. There
are $q + 1$ of the $\varepsilon_i$, which we take to be $x_1, x_2, ..., x_{q+1}$
in Eqs.~(\ref{eq:identity}). According to the first case in
Eqs.~(\ref{eq:identity}), the contribution to the last sum in Eq.~(\ref{f^2})
from $P(\varepsilon_i)$ vanishes, while according to the second case of
(\ref{eq:identity}), the contribution of $\varepsilon^{q}_i$ to the sum is
exactly 1. Thus, the components $f_n$ satisfy (\ref{absfn}). On the other hand,
it is easy to verify that if the $f_n$ satisfy (\ref{absfn}) then $b^{(i)}$
given by (\ref{eq:bj}) is an eigenvector with eigenvalue $\varepsilon_i$.

If $\varepsilon_0 < \alpha_0 < \varepsilon_1 < \ldots < \alpha_q$ then it can
easily be seen that the argument to the square root in Eq.~(\ref{absfn}) is
positive, and therefore a solution for the $f_n$ exists. Conversely, we can show
that if the argument to the square root is positive for all $n$, the eigenvalues
must have this ordering. For each $n=0,\ldots,q$ let $k_n$ be the number of
eigenvalues $\varepsilon_i$ for which $\varepsilon_i < \alpha_n$. Then
\begin{eqnarray*}
  \text{sign} \, \frac{\prod_{k = 0}^q (\alpha_n - \varepsilon_k)}
  {\prod_{k \neq n} (\alpha_n - \alpha_k)} & = & \frac{(-1)^{q+1-k_n}}{(-1)^{q-n}}.
\end{eqnarray*}
In order for the sign to be positive for all $n$, we therefore require that
$n+1-k_n$ is always even. But since $0 \leq k_0 \leq \ldots \leq k_q \leq q+1$,
the only possibility is $k_n = n+1$, i.e., $\varepsilon_0 < \alpha_0 <
\varepsilon_1 < \ldots < \alpha_q$.

The eigenvectors are given by Eq.~(\ref{eq:bj}). To determine the normalization constant $C_i$,
we require $\sum_{n=0}^q |b^{(i)}_n|^2 = 1$ and use Eq.~(\ref{absfn}) for $|f_n|$ to find
\begin{eqnarray*}
  C_i^{-2} & = & \sum_{n = 0}^q \frac{\prod_{0 \leqslant k
      \leqslant q} (\alpha_n - \varepsilon_k)}{\prod_{k \neq n, 0 \leqslant k
      \leqslant q} (\alpha_n - \alpha_k)}  \frac{1}{(\alpha_n -
    \varepsilon_i)^2}.
\end{eqnarray*}
This sum can also evaluated using the identity (\ref{eq:identity}). To do this,
we define $\alpha_{q+1} \equiv \varepsilon_i$ and rewrite
\begin{eqnarray*}
  C_i^{-2} & = & \sum_{n=0}^q \frac{\prod_{k \neq i, 0 \leqslant k \leqslant q} (\alpha_n - \varepsilon_k)}
  {\prod_{k \neq n, 0 \leqslant k \leqslant q+1}(\alpha_n - \alpha_k)}.
\end{eqnarray*}
We now add and subtract the $n=q+1$ term to obtain
\begin{eqnarray*}
  C_i^{-2} & = & \sum^{q + 1}_{n = 0} \frac{\prod_{k \neq i, 0 \leqslant k \leqslant
      q} (\alpha_n - \varepsilon_k)}{\prod_{k \neq n, 0 \leqslant k \leqslant q +
      1} (\alpha_n - \alpha_k)} \\
  &   & \qquad \qquad - \frac{\prod_{k \neq i, 0 \leqslant k \leqslant
      q} (\alpha_{q + 1} - \varepsilon_k)}{\prod_{0 \leqslant k \leqslant q}
    (\alpha_{q + 1} - \alpha_k)}.\\
\end{eqnarray*}
Taking $x_1=\alpha_0, x_2=\alpha_1, ..., x_N=\alpha_{q+1}$, the first case of
(\ref{eq:identity}) may be applied to find that the sum vanishes and thus
\begin{eqnarray*}
  C_i^{-2} & = & \frac{\prod_{k \neq i} (\varepsilon_k -
    \varepsilon_i)}{\prod_{k} (\alpha_k -
    \varepsilon_i)} \;.
\end{eqnarray*}
This completes the proof.

Since the eigenvalues $\varepsilon_i$, $\alpha_n$ for the cold atom problem are
indeed ordered as the theorem requires, we choose $f_n=|f_n|$ to be positive
real numbers according to Eq.~(\ref{absfn}).

The complete set of the two-particle eigenvectors of the
effective interaction $V_{\rm eff}^{(q)}$ is therefore given by (i) the noninteracting states
$|\mathcal{N}\mathcal{L}\mathcal{M}n l m \rangle$ for $l > 0$ or ($n > q$ and $l
= 0$) with eigenvalues $\left(2\mathcal{N} + \mathcal{L} + 3/2 + 2 n + l + 3 /
  2\right)\hbar\omega$, and (ii) the $l=0$ interacting states
\begin{multline}
  |\mathcal{N} \mathcal{L} \mathcal{M} b^{(i)} \rangle = \left [ \frac{\prod_{k} (\alpha_k -
      \varepsilon_i)}{\prod_{k \neq i} (\varepsilon_k -
      \varepsilon_i)} \right ] ^{1/2}\\
  \times \sum_{n=0}^q \frac{f_n}{\alpha_n - \varepsilon_i}
  |\mathcal{N}\mathcal{L}\mathcal{M}n 0 0 \rangle
\end{multline}
with eigenvalues $\left(2\mathcal{N} + \mathcal{L} + 3/2 +
  \varepsilon_i\right)\hbar\omega$ and where the $\varepsilon_i$ are given by
the solutions of Eq.~(\ref{eq:energies}).

We also derive closed expressions for the trace and norm of the
$(q+1)$-dimensional matrix $V_{\rm eff}^{(q)}$ defined by $(V_{\rm
  eff}^{(q)})_{n n'} \equiv \langle n 0 0|\hat{V}_{\rm eff}^{(q)}|n' 0
0\rangle$. Its trace has a simple form which can be obtained via
Eqs.~(\ref{absfn}) and (\ref{eq:identity}). Taking $x_1=\alpha_0, x_2=\alpha_1,
..., x_N=\alpha_{q}$ and applying the second and third cases of
Eq.~(\ref{eq:identity}), we find
\begin{eqnarray*}
  \text{tr}\ V_{\rm eff}^{(q)} & = & \sum_{n = 0}^q  \frac{\prod_k (\alpha_n - \varepsilon_k)}
  {\prod_{k \neq n} (\alpha_n - \alpha_k)} \\
  & = & \sum_{n=0}^q (\alpha_n - \varepsilon_n).
\end{eqnarray*}
This result implies that the Frobenius norm of $V_{\rm eff}^{(q)}$ is
\begin{eqnarray*}
  \sum_{n, n'} |(V_{\rm eff}^{(q)})_{n n'} |^2 =
  \left( \sum_{n = 0}^q (\alpha_n - \varepsilon_n) \right)^2\;.
\end{eqnarray*}

\subsection{Properties in the unitary limit}

In this section we consider the properties of the separable effective
interaction in the unitary limit. The simple form of the relative-motion unitary
eigenvalues $\varepsilon_i$ allows us to verify Eq.~(9) in
Ref.~\onlinecite{Alhassid:2008} for the interaction parameters $f_n$ at
unitarity and prove analytically that $V^{(q)}_{\rm eff}$ converges to the
pseudopotential (\ref{eq:pseudo}) in the limit $q \to \infty$.

\subsubsection{Interaction parameters $f_n$}

In the unitary limit $\varepsilon_i = 2 i + 1 / 2$ are the harmonic oscillator
energies shifted down by one oscillator quantum. In this case the expression for
$f_n$ reduces to
\begin{eqnarray}
  f_n^2 & = & \frac{\prod_{k = 0}^q (2 (n - k) + 1)}{\prod_{k\neq n} 2 (n - k)}
  \nonumber \\
  & = & \frac{\prod_{k = 0}^n (2 k + 1)}{\prod_{k = 1}^n 2 k}  \frac{\prod_{k =
      1}^{q - n} (- 2 k + 1)}{\prod_{k = 1}^{q - n} (- 2 k)} \nonumber \\
  & = & \frac{(2 n + 1)!!}{(2 n) !!}  \frac{(2 (q - n) - 1) !!}{(2 (q - n)) !!}\;,
  \label{eq:fnunitary}
\end{eqnarray}
which is exactly Eq.~(9) of Ref.~\onlinecite{Alhassid:2008}.

\subsubsection{Convergence of two-particle eigenvectors} \label{sec:convergence}

To prove the convergence of the effective interaction to the pseudopotential
$V({\bf r})$ in Eq.~(\ref{eq:pseudo}), we study the convergence of the
two-particle eigenstates $|b^{(i)} \rangle$ of $\hat{H}_0 + \hat{V}_{\rm
  eff}^{(q)}$ to the corresponding eigenstates $|u^{(i)}\rangle$ of $\hat{H}_0 +
V(r)$ as $q \rightarrow \infty$. For any scattering length, the square of the
$n$-th component of $|b^{(i)} \rangle$ is given by
\begin{eqnarray*}
  (b^{(i)}_n)^2 & = & \frac{\prod_k (\alpha_k - \varepsilon_i)}{\prod_{k \neq i}
    (\varepsilon_k - \varepsilon_i)}  \frac{\prod_k (\alpha_n -
    \varepsilon_k)}{\prod_{k \neq n} (\alpha_n - \alpha_k)}  \frac{1}{\left(
      \alpha_n - \varepsilon_i \right)^2} \;.
\end{eqnarray*}
In the unitary case
\begin{eqnarray*}
  (b^{(i)}_n)^2 & = & \frac{\prod_k (2 (k - i) + 1)}{\prod_{k \neq i} 2 (k -
    i)}  \frac{\prod_k (2 (n - k) + 1)}{\prod_{k \neq n} 2 (n - k)}
  \frac{1}{\left( \alpha_n - \varepsilon_i \right)^2}\\
  & = &  \frac{1}{(\alpha_n - \varepsilon_i)^2} \frac{(2 i - 1) !!}{(2 i) !!}
  \frac{(2 (q - i) + 1) !!}{(2 (q - i)) !!} \\
  &   & \times \frac{(2 n + 1) !!}{(2 n) !!} \frac{(2 (q - n) - 1) !!}{(2 (q - n)) !!} \;.
\end{eqnarray*}
The asymptotic behavior for large $q$ is
\begin{align*}
  b_n^{(i)} (q) \sim & \frac{\sqrt{2 / \pi}}{(\alpha_n - \varepsilon_i)}
  \left[ \frac{(2 i - 1) !!}{(2 i) !!} \right]^{1 / 2}  \left[ \frac{(2 n + 1)
      !!}{(2 n) !!} \right]^{1 / 2}  \\
  & \times \left[ \frac{q - i + 3 / 4}{q - n + 1 / 2}
  \right]^{1 / 4} \;.
\end{align*}
We compare to the exact wave function $| u^{(i)} \rangle$ by computing the
relative difference for each component,
\begin{eqnarray*}
  \frac{|b_n^{(i)} - u_n^{(i)} |}{u_n^{(i)}} & \sim & \left| \left( 1 + \frac{n
        - i + 1 / 4}{q - n + 1 / 2} \right)^{1 / 4} - 1 \right| \\
  & \sim & \frac{n - i + 1 / 4}{4 (q - n + 1 / 2)} \;.
\end{eqnarray*}
This diminishes like $1/q$ for large $q$, proving convergence to the exact
two-body eigenstates in relative motion. Importantly, this is faster than the
$1/\sqrt{q}$ convergence of the renormalized contact
interaction~\cite{Alhassid:2008}.

\section{Applications}

To demonstrate the advantage of the new effective interaction as compared with
the renormalized contact interaction, we study the convergence versus $q$ of
low-lying energies of the three and four-particles systems at various values of
the scattering length. We also study the convergence versus $q$ of the
three-particle free energy at unitarity (for which an exact solution exists).
The latter is important for thermodynamical studies of the trapped gas. The
three-particle system we study consists of two spin-up particles and one
spin-down particle ($\uparrow \uparrow \downarrow$), while the four-particle
system is unpolarized ($\uparrow \uparrow \downarrow \downarrow$).

\subsection{Spectroscopy}

The CI method works in the laboratory frame and requires a certain truncation
scheme. As in Ref.~\onlinecite{Alhassid:2008}, we use as our many-particle basis
the set of all Slater determinants which can be constructed from the
single-particle states in the harmonic oscillator shells $N=0,\ldots,N_{\rm
  max}$. For a given regularization parameter $q$, the two-body interaction
matrix elements in the laboratory frame are calculated by transforming to the
relative and center of mass coordinates via the Talmi-Moshinsky
brackets~\cite{Talmi:1952} and using the interaction (\ref{eq:effective}) in
relative motion. We carry out direct diagonalization of the CI Hamiltonian using
a new code for three and four particles which works in a two-species formalism
with a basis of good total orbital angular momentum $L$ and parity $\pi$. In
order to determine the many-particle energies, we take the following limits. For
fixed $q$ we calculate the energies as a function of $N_{\rm max}$ and
extrapolate to $N_{\rm max} \to \infty$ following the method of
Ref.~\onlinecite{Alhassid:2008}. The resulting energies are then studied as a
function of $q$ to obtain an estimate of the $q \to \infty$ limit.

In the CI calculations we used PARPACK for parallel diagonalization on a Linux
cluster using between 1 and 320 CPU cores. The largest system studied, the
four-particle system with $N_{\text{max}}=12$ and total $L=1$ (involving $\sim
4.6$ million configurations), required a total of about 2300 GB to store the
matrix elements of the Hamiltonian and 24 hours to run on 320 cores, while more
typical calculations, e.g., four particles at $N_{\text{max}}=9$ and $L=1$
($520,000$ configurations) required about 64 GB of storage and four hours of
computation time on 32 cores.

\subsubsection{Three particles}

We study a selected set of the low-lying energies of the three-particle system
as a function of the inverse scattering length. Here the main topic of interest
is the convergence of the eigenvalues (i) as the size of the model space (i.e.,
the maximal number of oscillator shells $N_{\rm max}$) increases, and (ii) as
the regularization parameter $q$ increases.

(i) Overall, convergence in $N_{\rm max}$ is fast for both the renormalized
contact interaction and the new effective interaction, and similar to what was
already discussed in Ref.~\onlinecite{Alhassid:2008} in the unitary limit.
Convergence slows somewhat as we follow the crossover from the BCS to the BEC
regime and tends to require more shells as $q$ increases. At $q=4$, $N_{\rm
  max}=13$, we find that the lowest four eigenvalues for each of $L^{\pi}=0^+$,
$1^+$, $1^-$, $2^+$, and $2^-$ are converged in $N_{\rm max}$ to about
$0.0003\%$ in the BCS regime ($a_{\rm osc}/a=-0.8$), $0.002\%$ in the unitary limit and
$0.01\%$ in the BEC regime ($a_{\rm osc}/a)=0.8$), on average.

\begin{figure}
  \begin{center}
    \includegraphics[width=0.5\textwidth]{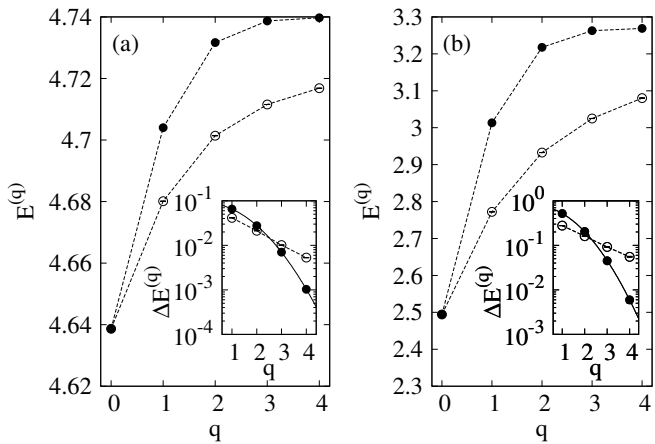}
    \caption{Convergence of the lowest 1- state for the three particle system on
      either side of the crossover. The energy $E^{(q)}$ is plotted as a
      function of the regularization parameter $q$. Solid circles are for the
      effective interaction and open circles are for the renormalized contact
      interaction. (a): The BCS side, $a_{\rm osc}/a = -0.8$. (b): The BEC side,
      $a_{\rm osc}/a=+0.8$. Inset: the differences $\Delta E^{(q)} \equiv
      E^{(q)} - E^{(q-1)}$ on a log scale. For the effective interaction the energy
      decreases slightly above $q=4$, with the $q=6$ values equal to
      $4.7392\,\hbar\omega$ and $3.2664\,\hbar\omega$ in cases (a) and (b), respectively.}
    \label{fig:A3_1-}
  \end{center}
\end{figure}

\begin{figure}
  \begin{center}
    \includegraphics[width=0.5\textwidth]{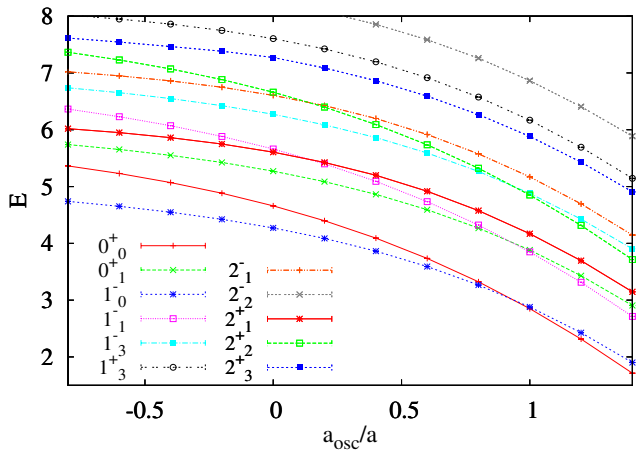}
    \caption{(Color online) Selected low-lying energies of the three-particle
      system vs. scattering length. Symbols are the estimates for $q \to \infty$
      as described in the text. Lines are linear interpolations.}
    \label{fig:A3evals}
  \end{center}
\end{figure}

(ii) As a function of $q$ (for $1 \leqslant q \leqslant 6$), we identify two
convergence patterns for low-lying three-body eigenvalues. The first occurs for
most eigenvalues, and is characterized by fast, monotonic convergence for $1
\leqslant q \leqslant 4$ or $1 \leqslant q \leqslant 5$ at an exponential or
faster-than-exponential rate. The values at larger $q$ then give small, often
nonmonotonic corrections. As an example, the lowest $L^{\pi}=0^+$ state at
unitarity has the exact value $E=4.6662\ldots\hbar\omega$. With the effective
interaction, we find the energy of this state decreases monotonically to
$E=4.6602\, \hbar \omega$ at $q=4$ (slightly overshooting), and
$E=4.6597\,\hbar\omega$ at $q=5$. The convergence then switches direction at
$q=6$ and approaches the exact value with a low power-law behavior, so that the
$q \geqslant 6$ values provide only a $0.14\%$ correction to the $q=4$ value.
For such eigenvalues we report the energy at the highest $q$ calculated, usually
$q=6$.

The second convergence pattern we encounter is completely monotonic for $2
\leqslant q \leqslant 6$, but with a noticeably slower rate of convergence. For
these eigenvalues, the convergence is smooth, and we estimate the $q \to \infty$
value by fitting ${\rm log}(\Delta E^{(q)})$, where $\Delta E^{(q)} = E^{(q)} -
E^{(q-1)}$, to a quadratic polynomial. This provides an interpolating function
$f(q) = A + B q + C q^2$ (with $C<0$) which we then use to numerically sum the
residual terms in the expansion $E^{(\infty)} = E^{(q)} + \Delta E^{(q+1)} +
\Delta E^{(q+2)} + \ldots \approx E^{(q)} + 10^{f(q+1)} + 10^{f(q+2)} + \ldots$.
An example of such a state is the $0^+_3$ state at $a_{\rm osc}/a=0.0$. (We
label energy levels as $L^{\pi}_n$, where $n$ indexes the levels with total
angular momentum $L$ and parity $\pi$ in increasing order starting from zero, so
that $0^+_3$ is the fourth-largest eigenvalue of those with $L=3$ and positive
parity.) In this case, the $q=4$ value is $6.716\,\hbar\omega$ and the
extrapolated value (from $q=6$) is $6.657$, indicating the state with the exact
energy of $6.6662\ldots\hbar\omega$, a $0.14\%$ error.

Comparing with the renormalized contact interaction, we find significantly
faster convergence with the new interaction for both convergence patterns. For
example, the $0^+_3$ state at unitarity attains the value $6.984\,\hbar\omega$
at $q=4$ with the contact interaction, a difference of $5\%$ from the correct
value, compared to $0.7\%$ at $q=4$ for the new interaction. The difference for
faster-converging eigenvalues is not as dramatic, but the new interaction is
still clearly preferred. This is shown in Fig.~\ref{fig:A3_1-} for the $1^-_0$
state on either side of the crossover, at $a_{\rm osc}/a=-0.8$ and $a_{\rm
  osc}/a=0.8$. Importantly, we see that the advantage of the new interaction is
maintained away from unitarity. This holds true for all the eigenvalues we
studied, and for the four-particle system as well.

Because convergence is only truly attained for large $q$, and it is difficult to
assess the effects of non-monotonicity, we estimate the systematic error in our
three-particle energies along the crossover based on comparison with the unitary
calculations, where we compare the convergence in $q$ with the known exact
results. From this we estimate an upper bound for the systematic error due to
non-convergence of approximately $0.2\%$ in our three-particle energies.

Using the new effective interaction, we also present in Fig.~\ref{fig:A3evals}
results for a selected set of the low-lying energies of the three-particle
system as a function of the inverse scattering length. Although the
three-particle system is now well-studied, this provides a clear verification of
the capability of the method. Our results agree with those published
previously~\cite{Stetcu:2008,Kestner:2007}. As in the the earlier studies, we
find crossings between levels with different quantum numbers, for instance the
$2^{+}_1$ state and the $1^{-}_1$ state at approximately $a_{\rm osc}/a = 0.14$,
and the $1^-_0$ state and the $0^+_0$ state at approximately $a_{\rm osc}/a =
0.93$, which changes the parity and angular momentum of the ground state.

\subsubsection{Four particles}

The four-particle system is less well-studied and provides a more interesting
test of the new effective interaction. As with the three-particle system, we
examined a selected set of the low-lying energies as a function of the inverse
scattering length.

(i) Overall we find fast convergence in $N_{\rm max}$, very similar to the
three-particle case. More precisely, at $q=4$, $N_{\rm max}=11$, the level of
convergence ranges from an average of about $0.05\%$ at $a_{\rm osc}/a=-0.8$ to
$0.2\%$ at $a_{\rm osc}/a=0.8$ for the lowest four states of each of the $0^-$,
$0^+$, $1^-$ and $1^+$ configurations.

(ii) As a function of $q$, the two convergence patterns observed for the three
particle case -- the more common fast, slightly non-monotonic convergence and
the less common slower, monotonic convergence -- are still prevalent, but with
the addition of a few eigenvalues which are non-monotonic for smaller $q$. An
example is the $0^{+}_1$ state at unitarity, which becomes non-monotonic at
$q=3$. For eigenvalues with the first two convergence patterns, we estimate the
$q \to \infty$ values in the same manner as the three-particle case. For the new
non-monotonic eigenvalues, we report the result of the calculation at largest
$q$, usually $q=5$. For instance, for the $0^{+}_1$ state at unitarity, the
$q=5$ value is $E=7.036\,\hbar\omega$, and appears to be an upper bound based on
the direction of convergence. This compares favorably with the result
$E=7.010\,\hbar\omega$ (a $0.4\%$ difference) of Ref.~\onlinecite{daily:2010},
in which the center of mass coordinate is separated out and a stochastic
variational approach is employed. Note we add a center of mass excitation of
$1.5\,\hbar\omega$ to the results of Ref.~\onlinecite{daily:2010} to allow
comparison.

An example of a fast, slightly non-monotonic energy is the $1^+_0$ state, which
was calculated in Ref.~\onlinecite{daily:2010} to have a value
$E=6.588(20)\,\hbar\omega$. Our $q=5$ estimate of the new interaction is
$6.582\,\hbar\omega$, well within error.

\begin{figure}
  \begin{center}
    \includegraphics[type=eps,ext=.eps,read=.eps,width=0.5\textwidth]{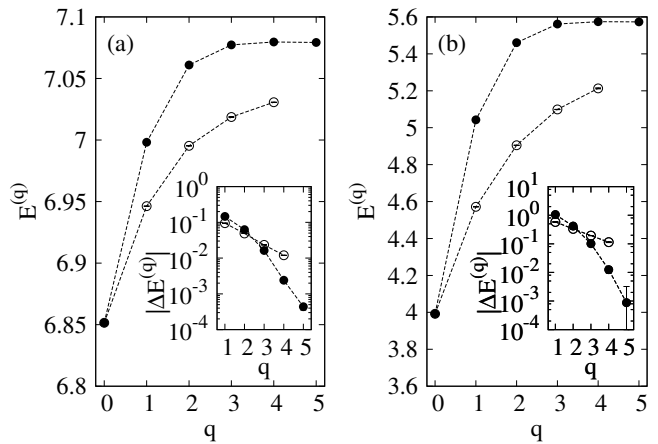}
    \caption{Convergence of the lowest $1^{+}$ state for the four particle
      system. Solid circles are the results calculated with the effective
      interaction, and open circles are the results from the renormalized
      contact interaction. (a): Convergence of the energy $E^{(q)}$ vs $q$ on
      the BCS side of the crossover at $a_{\rm osc}/a=-0.8$. (b): The BEC side of the
      crossover at $a_{\rm osc}/a=+0.8$. Inset: The absolute values of the differences
      $|\Delta E^{(q)}|$. In both cases, the convergence becomes slightly
      non-monotonic at the $q=5$ point.}
    \label{A4_1+}
  \end{center}
\end{figure}

Compared with the renormalized contact interaction, the new interaction provides
much improved convergence for four particles, as it did for the three particle
system. For example, Fig.~\ref{A4_1+} shows the convergence of the $1^{+}_0$
state using the new and the renormalized contact interactions on either side of
the crossover. As with all the energies we have studied, this displays a marked
improvement in convergence. On the BEC side, in particular, we find a $7\%$
difference between the two interactions at $q=4$.

\begin{figure}
  \begin{center}
    \includegraphics[width=0.45\textwidth]{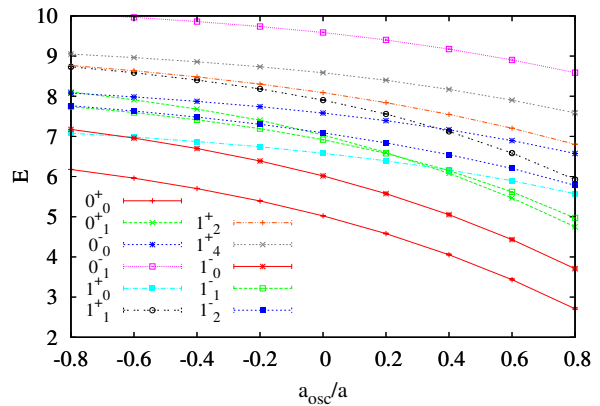}
    \caption{(Color online) Selected low-lying energies of the four-particle
      system vs. scattering length. Symbols are the estimates for $q \to \infty$
      as described in the text. Lines are linear interpolations.}
    \label{fig:A4evals}
  \end{center}
\end{figure}

In Fig.~\ref{fig:A4evals} we present results for a selected set of low-lying
energies of the four-particle system as a function of inverse scattering length.
As for the three-particle case, we find crossings between levels with different
quantum numbers, for instance, the $1^{+}_1$ state and the $0^{-}_0$ state at
approximately $a_{\rm osc}/a = 0.36$. In contrast to the three-particle system,
the angular momentum and parity of the ground state (here $0^{+}_0$) remain
fixed through the full range of scattering lengths.

\subsection{Thermodynamics}

As thermodynamics is one of the central topics of interest for cold atoms, we
test the usefulness of the new separable effective interaction for
thermodynamical studies by studying the convergence of the free energy (at a
given temperature) versus $q$ for the three-particle system. Our calculations
were done at temperatures between $T=0\,\hbar\omega$ and $T=2.0\,\hbar\omega$ as
a function of $q$ and $N_{\rm max}$ via a direct diagonalization of the
Hamiltonian through $N_{\rm max}=10$. We compare our calculations with exact
results for the three-particle system.

\begin{figure}
  \begin{center}
    \includegraphics[type=eps,ext=.eps,read=.eps,width=0.5\textwidth]{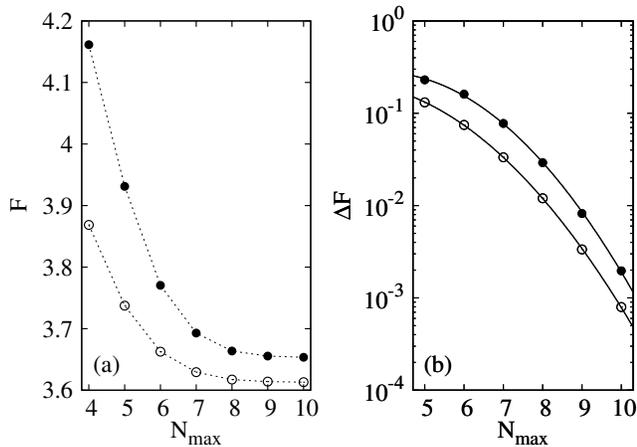}
    \caption{Convergence of the free energy in $N_{\rm max}$ at $T=0.4 \hbar
      \omega$ and $q=4$ calculated with the effective interaction (solid
      circles) and renormalized contact interactions (open circles). The solid
      lines are fits to the interpolating function ${\rm log}(\Delta F)=A + Bq +
      Cq^2$ with fit parameters $A$,$B$,$C$, and are used to extrapolate to the
      limit $N_{\rm max}~\to~\infty$.}
    \label{fig:FvsNmax}
  \end{center}
\end{figure}

An example of the convergence in $N_{\rm max}$ for $q=4$ and $T=0.4\,\hbar
\omega$ is shown in Fig.~\ref{fig:FvsNmax}, where we observe smooth behavior for
both interactions. For higher temperatures, the smooth character of the $\Delta
F$ vs. $N_{\rm max}$ is preserved, while the curvature flattens out to a
straight line (on a logarithmic scale), allowing for accurate extrapolations in
$N_{\rm max}$.

\begin{figure}
  \begin{center}
    \includegraphics[type=eps,ext=.eps,read=.eps,width=0.48\textwidth]{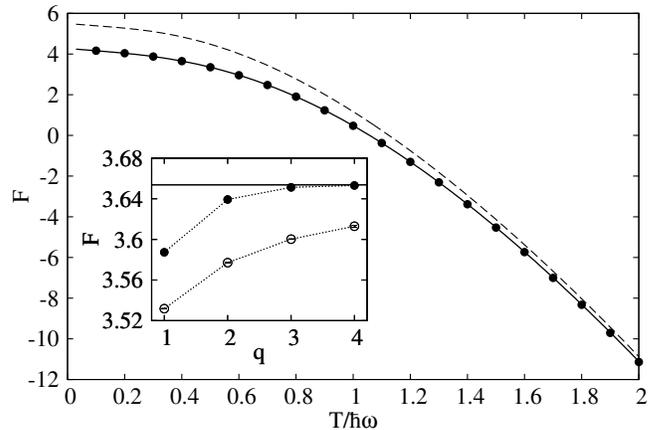}
    \caption{Free energy versus temperature for the three-particle system.
      Solid circles: new effective interaction at $q=4$. Solid line: exact result. Dashed
      line: non-interacting system. The inset demonstrates the convergence of $F$ vs. $q$ at
      $T=0.4\,\hbar\omega$. The solid circles were calculated using the separable effective
      interaction, while the open circles are from the renormalized contact
      interaction. The horizontal line indicates the exact result of
      $3.6537\ldots\hbar\omega$.}
    \label{fig:FvsT}
  \end{center}
\end{figure}

The convergence as a function of $q$ is demonstrated in the inset to
Fig.~\ref{fig:FvsT} for $T=0.4\;\hbar \omega$. The convergence using the new
effective interaction is significantly faster than for the renormalized contact
interaction. In general, we found that the convergence in $q$ for the new
interaction was quite uniform and fast for temperatures up to approximately
$T=1.0\,\hbar\omega$. In fact, at higher temperatures, the abundance of
interactionless states (see below) diminishes the error for even small $q$
values so that at $T=2.0\,\hbar\omega$, the difference between $q=1$ and the
exact value is only about $0.8\%$, compared to $2.1\%$ at $T=0.1\,\hbar\omega$.
By comparison, the difference in the free energy between the non-interacting and
interacting systems is $2\%$ at $T=2.0\,\hbar\omega$ and $30\%$ at
$T=0.1\,\hbar\omega$.

\begin{figure}
  \begin{center}
    \includegraphics[type=eps,ext=.eps,read=.eps,width=0.48\textwidth]{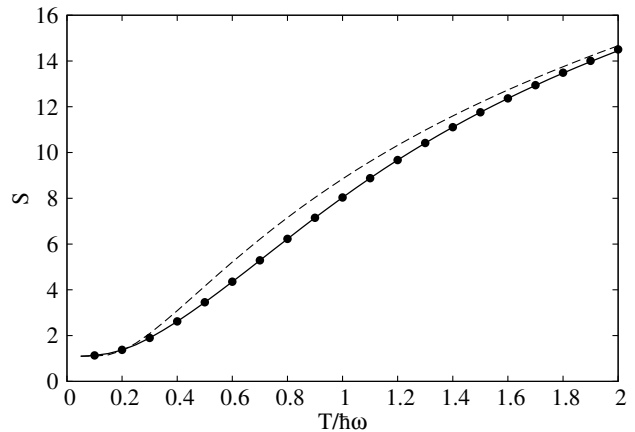}
    \caption{Entropy versus temperature for the three-particle system,
      calculated via a numerical derivative of the free energy, $S=-\partial
      F/\partial T$. Symbols and lines are as in Fig.~\ref{fig:FvsT}.}
    \label{fig:SvsT}
  \end{center}
\end{figure}

Our thermodynamics studies are summarized in Figs.~\ref{fig:FvsT} and
\ref{fig:SvsT}, where we compare our calculations of the free energy and entropy
as a function of temperature with exact calculations for the interacting and
non-interacting systems (described below). We find excellent agreement. Although
this is only a three-body system, the entropy curve indicates that our results
apply in the entirety of the interesting quantum degenerate regime.

Since the three-body problem has been solved exactly at
unitarity~\cite{werner:2006,tan3body}, we can compare the thermodynamic
functions calculated using the new effective interaction with exact results.

We summarize the exact calculation as follows. There are two classes of
eigenstates for the trapped Fermi gas, \emph{interacting} states and
\emph{interactionless} states. The interacting states satisfy the Bethe-Peierls
contact condition~\cite{werner:2006}
\[ \varphi(\bm{r}_1,\bm{r}_2,\bm{r}_3) = \left( \frac{1}{r_{ij}} - \frac{1}{a}
\right) A(\bm{R}_{ij},\bm{r}_k) + O(r_{ij}) \] (as $r_{ij} \rightarrow 0$) via a
function $A$ which is non-zero. The interactionless states satisfy a similar
condition but with $A$ identically zero. They are also eigenstates of the
non-interacting Hamiltonian and are identical for all values of the scattering
length.

To calculate the energy levels of the interacting states we follow the procedure
of Ref.~\onlinecite{werner:2006}. However, as noted there, at large temperatures
the number of interactionless states grows much larger than that of the
interacting states, so that energies for both are necessary to calculate
thermodynamic quantities for the system. The wave functions and energies of the
interactionless states for the three-particle system have been characterized
completely in Ref.~\onlinecite{tan3body}. Below we summarize the main results.

The energy of any state of the system (interacting or interactionless) is given
by
\[ E = (\gamma + n_{c} + 2 \nu + 9/2)\hbar \omega \] where $n_c$ is the number
of oscillator quanta in the center-of-mass excitation, $\gamma$ is a scaling
exponent, and $\nu = 0,1,2,...$ is the hyperradial quantum number. The
corresponding intrinsic wave functions are of the form $\psi \propto
\tilde{\phi} \, L_\nu^{(\gamma+2)}(R^2/a_{\rm osc}^2) e^{(-R^2/2a_{\rm
    osc}^2)}$, where
$R$ is the hyperradius. The function $\tilde{\phi}$ ($\propto R^\gamma$)
satisfies the Bethe-Peierls contact conditions as well as additional equations.
States with total internal angular momentum $l=0,1,\ldots$ have exponents
$\gamma_{l,n}$ where $n=0,1,\ldots$. For the interacting states
$\gamma_{l,n}=s_{l,n} -7/2$ where $s_{l,n}$ are an infinite sequence of
solutions to an $l$-dependent transcendental equation obtained from the
three-particle Schrodinger equation~\cite{werner:2006}. For the interactionless
states, $\gamma_{l,n}$ are integers $\geq 2$ and have degeneracies; these
degeneracies are non-trivial and must be determined in order to calculate
thermodynamic quantities for the system. We followed the method of
Ref.~\onlinecite{tan3body} to determine these degeneracies (see, e.g., Table II
of Ref.~\onlinecite{tan3body}).

Using the above classification of interacting and interactionless states, we
computed the entire three-body spectrum to the extent needed to ensure
convergence of the free energy at any temperature of interest (see
Figs.~\ref{fig:FvsT} and \ref{fig:SvsT}). Table \ref{tab:FSvsT} provides a few
values of the free energy $F$ and entropy $S$ at various temperatures for
reference. For the non-interacting calculations we use simple closed formulas
that are derived in the Appendix.

\begin{table}
  \begin{ruledtabular}
    \begin{tabular}{ccccccc}
      $T \,\, (\hbar \omega)$ &  $E \,\,(\hbar \omega)$ &  $F \,\,{\bf (\hbar \omega)}$ &
      $S$ & $E_{\text{free}}$  &  $F_{\text{free}}$  &   $S_{\text{free}}$   \\
      \hline
      0.1  & 4.2754 & 4.1622    & 1.1321 & 5.5003 & 5.3901   & 1.1016  \\
      0.5  & 5.0767 & 3.3510    & 3.4514 & 6.5632 & 4.4837   & 4.1590  \\
      1.0  & 8.5107 & 0.47585   & 8.0348 & 10.017 & 1.1690   & 8.8483  \\
      2.0  & 17.792 & -11.109   & 14.451 & 18.459 & -10.874  & 14.667  \\
    \end{tabular}
  \end{ruledtabular}
  \caption{\label{tab:FSvsT}A few values of the exact thermodynamic functions of the
    unitary and non-interacting three-particle systems. $E$, $F$, and $S$ are
    the energy, free energy and entropy of the unitary system;
    $E_{\text{free}}$, $F_{\text{free}}$, and $S_{\text{free}}$ are the same
    quantities for the non-interacting system.}
\end{table}

\section{Conclusion}

We have extended a new separable effective interaction for the unitary trapped
gas in the configuration-interaction framework to regions away from unitarity,
i.e., the BEC-BCS crossover. The main advantage of this interaction, when
compared with the conventional renormalized contact interaction, is its much
improved convergence of the many-particle energies in the
regularization parameter. This allows accurate calculations in the configuration
interaction approach. In particular, we calculated the low-lying energies for
three and four particles as a function of the inverse scattering length.

We also studied the three-particle system at finite temperature in the unitary
limit and found similar convergence properties of the free energy as a function
of the regularization parameter. This separable effective interaction may
therefore facilitate accurate studies of the thermodynamics of the trapped cold
atom system.

\begin{acknowledgments}

  We acknowledge S. Tan for providing us with Ref.~\onlinecite{tan3body} and G.F.
  Bertsch for useful discussions. We also thank I. Stetcu for sharing with us
  certain three-body calculations using the new interaction. This work was
  supported in part by the U.S. DOE grant No. DE-FG02-91ER40608, by facilities
  and staff of the Yale University Faculty of Arts and Sciences High Performance
  Computing Center, and by the NSF grant No. CNS 08-21132 that partially funded
  acquisition of the facilities. Computational cycles were also provided by the
  NERSC high performance computing facility at LBL.

\end{acknowledgments}

\appendix
\section{Non-interacting thermodynamics}

In Figs.~\ref{fig:FvsT} and \ref{fig:SvsT} we compared our calculations with the
non-interacting three-body trapped systems (dashed lines). For this purpose we
derived simple formulas for the canonical partition functions of small systems
of trapped non-interacting spin-$1/2$ particles. For three particles ($\uparrow
\uparrow \downarrow$), we proceed as follows. Label the state of each particle
as $\bm{s}=(n_x,n_y,n_z,\sigma)$, where $n_x,n_y,n_z$ are the $3$-D harmonic
oscillator quantum numbers and $\sigma=\pm 1/2$ is the $z$-component of the
spin. Then the canonical partition function is
\[ \mathcal{Z} = \frac{1}{3!}
\sum_{\substack{\bm{s}_1,\bm{s}_2,\bm{s}_3 \\
    \sigma_1 + \sigma_2 + \sigma_3 = 1/2}}
e^{-\beta \left( E_{\bm{s}_1} +
    E_{\bm{s}_2} + E_{\bm{s}_3} \right)} S(\bm{s}_1,\bm{s}_2,\bm{s}_3) \]
where $E_{\bm{s}} = (n_x + n_y + n_z + 3/2)\hbar\omega$ and
$S(\bm{s}_1,\bm{s}_2,\bm{s}_3)$ is the selection function
\begin{align}
  S(\bm{s}_1,\bm{s}_2,\bm{s}_3) & = \left\{
    \begin{array}{l l}
      0 & \quad \mbox{if any two $\bm{s}_i$ are equal}\\
      1 & \quad \mbox{otherwise}\\ \end{array} \right. \\
  & = \prod_{i<j} (1-\delta_{\bm{s}_i,\bm{s}_j}). \label{eq:selection}
\end{align}

To perform the sum, one expands the selection function to obtain eight terms
involving sums over the Maxwell-Boltzmann factor $e^{-\beta \left( E_{\bm{s}_1}
    + E_{\bm{s}_2} + E_{\bm{s}_3} \right)}$ times a product (say $D$) of zero to
three Kronecker $\delta$'s. By collapsing the sum over the $\delta$'s one
obtains a geometric series which then easily yields a closed-form expression.
As an example, for $D=\delta_{\bm{s}_1,\bm{s}_2}$, and with the notation $|\bm{n}_i|
\equiv n_{i,x}+n_{i,y}+n_{i,z}$, the corresponding sum is
\begin{equation*}
  \begin{split}
    \sum_{\substack{\bm{s}_1,\bm{s}_2,\bm{s}_3 \\
        \sigma_1 + \sigma_2 + \sigma_3 = 1/2}}
    &e^{-\beta \left( E_{\bm{s}_1} +
        E_{\bm{s}_2} + E_{\bm{s}_3} \right)} \delta_{\bm{s}_1,\bm{s}_2} \\
    &= \left( \frac{1}{1-e^{-2\beta}} \right)^3 \left( \frac{1}{1-e^{-\beta}}
    \right)^3 .
  \end{split}
\end{equation*}
Note that in this case, the sum over $\sigma_2$ and $\sigma_3$ is exactly 1 (and
is zero for the two and three-$\delta$ terms for this system).

The final result is
\begin{multline*}
  \lefteqn{\mathcal{Z} = \frac{1}{2} e^{-9\beta/2} \Bigg[
    \left(\frac{1}{1-e^{-\beta}}\right)^9 - }\\
  \left(\frac{1}{1-e^{-\beta}}\right)^3
  \left. \left(\frac{1}{1-e^{-2\beta}}\right)^3 \right].
\end{multline*}

\end{document}